\documentclass[aps,superscriptaddress,showpacs,nofootinbib]{revtex4}

\usepackage{amssymb}
\usepackage{epsf,epsfig}

\begin{document}

\title{Incorporating Memory Effects in Phase Separation Processes}

\author{T. Koide}
\email{koide@if.ufrj.br}
\affiliation{Instituto de F\'{\i}sica, Universidade Federal do Rio
de Janeiro, 21941-972 Rio de Janeiro, RJ,
Brazil}

\author{G. Krein}
\email{gkrein@ift.unesp.br}
\affiliation{Instituto de F\'{\i}sica Te\'orica,
Universidade Estadual Paulista,
01405-900 S\~ao Paulo, SP, Brazil}

\author{Rudnei O. Ramos}
\email{rudnei@uerj.br}
\affiliation{Departamento de F\'{\i}sica Te\'orica, Universidade do Estado do
Rio de Janeiro, 20550-013 Rio de Janeiro, RJ, Brazil}

\begin{abstract}
We consider the modification of the Cahn-Hilliard equation when a time delay
process through a memory function is taken into account. We then study the
process of spinodal decomposition in fast phase transitions associated with
a conserved order parameter. {}Finite-time memory effects are seen to affect
the dynamics of phase transition at short times and have the effect of
delaying, in a significant way, the process of rapid growth of the order
parameter that follows a quench into the spinodal region. These effects are
important in several systems characterized by fast processes, like
nonequilibrium dynamics in the early universe and in relativistic heavy-ion
collisions.

\medskip

\medskip

\noindent
keywords: nonequilibrium field dynamics, memory effects, 
relativistic heavy-ion collisions

\end{abstract}


\maketitle

The dynamics of phase transitions depends on whether the order
parameter that characterizes the different phases of a system is a
conserved quantity or not. In different fields of physics and
chemistry the dynamics of a conserved order parameter has been
described phenomenologically by the Cahn-Hilliard (CH)
equation~\cite{CH} (see also Ref.~\cite{bray-gunton} for
a review). However, it is a first order equation in time and as
such does not take into account memory effects (ME), which may have
quite important consequences on non-equilibrium dynamics for fast phase
transitions.

The CH equation is a diffusion-reaction type of
equation. 
Diffuse processes are characterized by microscopic
scattering events. 
The diffusion equation, in particular, 
describes the limiting situation where the time between the scattering 
events is 
infinitesimally short and therefore does not respect
causality. Since in real systems scattering events proceed through
finite time intervals, their description through diffusion-type of
equations poses serious problems in real physical situations, as
already recognized for a very long time~\cite{Jou1}.
Another related problem with the usual diffusion equation is that
it also leads to the breaking of the f-sum rule 
which gives the frequency sum of the dynamic structure factor
\cite{Kadanoff,Koidediff}.
However, when the time scales of the ME are much
smaller than any other scales, the breaking of causality and its
effects become negligible, like in several applications with the
CH equation, particularly in problems of metallurgy
(see for example the references in~\cite{Cahn}).

However, memory and causality constrains cannot be ignored when
the typical microscopic time scales are large in comparison with
the other time scales characterizing the dynamics. {}Fast phase
transitions are expected to have happened in the early universe
and most certainly also characterize the phase transitions
expected to occur in the highly excited matter formed in
relativistic heavy-ion collisions (RHIC). In the early universe
such situations may have happened when the typical microscopic
time scales for relaxation, given by the inverse of the decay
width associated with particle dynamics, is larger than the Hubble
time. This is a situation likely to be expected when
describing GUT phase transitions or even the inflationary dynamics
\cite{Rudnei}. In RHIC one expects to learn about the QCD phase
transition~\cite{Greiner,KoideM}. {}For instance, from the
hydrodynamic analysis of the freeze-out temperature in the most
central Au-Au collisions at~130~A~GeV, the typical reaction time
is around 10-20 fm/c~\cite{Hama}. However, the characteristic time
scale of the memory function in the Langevin equation which
describes the dynamics near the chiral phase transition is
predicted to be about 1~fm/c (as, for example, it is shown in {}Fig.~4
of Ref.~\cite{KoideM}), which is not short enough to be ignored.
These are examples that, when analyzing e.g. the detailed dynamics
related to conserved charges, may require the use of a related CH
equation for these conserved charges that goes beyond the linear
order in the time derivative and where ME can be accounted for.

In order to motivate causality in the CH equation, it is useful to
start our discussion by briefly reviewing the introduction of ME
in the diffusion equation. Consider a random walk process,
described by the simplest Langevin equation, $ \dot{x}(t) =
\xi(t)$, where $\xi (t)$ represents a noise term with correlations
$\langle \xi(t) \rangle = 0$ and $\langle \xi(t)\xi(t') \rangle =
2\Gamma W(t-t')$, with the intensity of the noise $\Gamma$. It is
common to ignore ME and assume a Gaussian white noise, for which
$W(t-t') = \delta(t-t')$. Then, the {}corresponding Fokker-Planck
equation for the probability distribution $P(x,t)$ is the usual
diffusion equation. However, if one takes ME into account through
a colored noise, for example of the form $W(t-t') =
\exp(-|t-t'|/\gamma)/\gamma$, one obtains the so-called causal
diffusion equation, which can be approximated by the partial
differential equation
\begin{eqnarray}
\gamma \frac{\partial^2}{\partial t^2}P(x,t) +
\frac{\partial}{\partial t}P(x,t) = \Gamma \, \nabla^2 P(x,t).
\label{diffusion}
\end{eqnarray}
The propagation speed of the equation is defined by
$\sqrt{\Gamma/\gamma}$ and one can easily see that the propagation
speed of the ordinary diffusion equation ($\gamma = 0$) is
infinite. A causal diffusion equation like Eq.~(\ref{diffusion})
has been used recently~\cite{gavin} to discuss the evolution of
conserved charges in RHIC. In the present paper we apply these
ideas to dynamic phase transitions involving a conserved order
parameter by deriving an analogous CH equation using memory
functions. To the best of our knowledge, this is the first
numerical analysis to incorporate memory constraints in a CH
equation.

Though our primary motivations are fast phase transitions in
cosmology and heavy-ion collision dynamics, we here adopt a very
general approach that can also be of relevance in other
applications, like in condensed matter systems. We shall then consider
the following general Ginzburg-Landau (GL) free energy
\begin{eqnarray}
{}F (\phi) = \int d^3 {\bf x} \left[ \frac{a}{2} (\nabla \phi)^2 -
\frac{b}{2} \phi^2 + \frac{c}{4} \phi^4 \right],
\label{F}
\end{eqnarray}
where $\phi$ is a conserved order parameter. To describe the phase
transition, we set, as usual, $b \sim T_c - T $ and hence the GL
free energy has two minima below $T_c$. Because the order parameter is
conserved it satisfies the equation of continuity
$\frac{\partial}{\partial t}\phi ({\bf x},t) + \nabla \cdot
{\bf J}({\bf x},t) = 0$,
where ${\bf J}({\bf x},t)$ is a current. The ordinary CH equation
follows assuming the irreversible current in the form
${\bf J}({\bf x},t) = - \Gamma \nabla  
\frac{\delta F(\phi)}{\delta \phi},$
where $\Gamma$ denotes here a kind of Onsager coefficient. Note that
the irreversible current is instantaneously produced by the
thermodynamic force $\nabla \frac{\delta F(\phi)}{\delta \phi}$
and as such time delay effects are not contained in this
formulation. However, the instantaneous assumption leads to
results in contradiction with those obtained in experiments, 
like in the
problems of heat conduction problem, spin diffusion, dielectric
relaxation, and so on (see discussions in respect to this in
Ref.~\cite{Jou1}). It is further known that sum rules are broken
under the instantaneous assumption, as discussed in
Ref.~\cite{Kadanoff,Koidediff}. To overcome these difficulties,
time delay or ME should be taken into account.

A traditional way \cite{Jou1,Kadanoff} to take into account ME is
to generalize the current defined above as follows,
\begin{eqnarray}
{\bf J}({\bf x},t)
= \! - \!\! \int^{t}_{0} \! ds \, d^3{\bf x}' {\cal M}({\bf x}-{\bf
x}',t-s) \nabla_{{\bf x}'} \frac{\delta F(\phi)}{\delta
\phi({\bf x'},s)} , \label{eqn:GFick}
\end{eqnarray}
where ${\cal M}({\bf x},t)$ is a memory function expressed by
the correlation function of noise, as required by the
fluctuation-dissipation theorem of second kind. {}Following the
experience with the diffusion equation, we use a local function in
space as
${\cal M} ({\bf x},t) = \Gamma \, e^{-t/\gamma}
\delta^{(3)}({\bf x})/\gamma$,
where $\gamma$ is the relaxation time of the memory function.
Substituting this into Eq.~(\ref{eqn:GFick}), we obtain
\begin{eqnarray}
\gamma \frac{\partial}{\partial t}{\bf J}({\bf x},t) = -{\bf J}({\bf
x},t) - \Gamma \nabla  \frac{\delta F(\phi)}{\delta \phi} \;,
\end{eqnarray}
which is the analogous to the
Maxwell-Cattaneo-type equation used in the heat conduction problem
\cite{Jou1}.
When substituting this equation into the equation of continuity,
we obtain the modified CH equation,
\begin{eqnarray}
\gamma \frac{\partial^2}{\partial t^2} \phi ({\bf x},t) +
\frac{\partial}{\partial t}\phi ({\bf x},t) = \Gamma \, \nabla^2
\frac{\delta F(\phi)}{\delta \phi} .
\label{causalCH}
\end{eqnarray}
When $\gamma = 0$, this equation reduces to the ordinary CH
equation. Eq.~(\ref{causalCH}) is our main result and in the
following we examine the practical consequences of incorporating
ME in the CH equation. We will also soon define precisely what we
mean by Eq. (\ref{causalCH}) to be causal, which will then imply a
constraint condition on $\Gamma$ and $\gamma$.

A simple way to assess the consequences of introducing memory in
the CH equation is to analyze the short-time dynamics of spinodal
decomposition (SD), {\it i.e.}, the process of phase separation following 
a quench into the two phase region of the phase diagram \cite{bray-gunton}. 
Recall that under these
circumstances that spinodal
decomposition is characterized by the exponential growth of long-wavelength
fluctuations in the order parameter at short times after the quench, 
leading to the formation of domains and coarsening at later times.
This is in contrast to the other possible mechanism for phase separation,
{\it i.e.} nucleation, where the process of phase transition initiates through
the decay of a metastable state by formation of bubbles of the stable 
phase that grow and percolates.
 
By making a linear approximation of
Eq.~(\ref{causalCH}), valid for small amplitude initial
conditions, we obtain the equation for the {}Fourier-transformed
field $\phi({\bf k},t)$,
\begin{eqnarray}
\gamma \frac{\partial^2}{\partial t^2} \phi ({\bf k},t)\! +\!
\frac{\partial } {\partial t}\phi ({\bf k},t) \approx -\Gamma
{\bf k}^2( a {\bf k}^2\! -\! b  )\phi ({\bf k},t).
\label{shorttime}
\end{eqnarray}
Taking an initial condition with zero time derivative and
$\phi({\bf k},t=0)=\phi_0({\bf k})$, the solution of this
equation can be written as
\begin{eqnarray}
\phi_c ({\bf k},t) = \phi_0({\bf k}) e^{-t/2\gamma}  ( \lambda_+
e^{\lambda \, t/2\gamma} - \lambda_- e^{-\lambda \, t/2\gamma}
)/(2 \lambda), \label{phi-c}
\end{eqnarray}
where $\lambda_{\pm} = 1 \pm \lambda$ and $\lambda = \sqrt{1 - 4
\gamma \Gamma {\bf k}^2 (a {\bf k}^2 -b )}$. On~the other hand,
the solution of the noncausal equation is
\begin{eqnarray}
\phi_{nc} ({\bf k},t) = \phi_0({\bf k})\, e^{-\Gamma{\bf k}^2 (a
{\bf k}^2-b) ) t} . \label{phi-nc}
\end{eqnarray}
The sub-indexes $c$ and $nc$ stand for causal and noncausal. The
long wavelength instability characterizing the SD happens for
wave-numbers such that the exponentials are larger than zero. This
happens for modes with momentum ${\bf k}^2 < b/a$, for both the
causal and noncausal solutions. {}For higher values of momentum,
$\phi_c ({\bf k},t)$ exhibits oscillation with relaxation, where
the relaxation time is given by the momentum independent constant
$2\gamma$, while $\phi_{nc}({\bf k},t)$ shows only relaxation
(decay), with a rate that increases infinitely with momentum.

Similarly, we can derive an approximate
solution at late times. Below the critical temperature the
parameter $b$ is positive and the order parameter condenses with
the size $\sqrt{b/c}$. We can then expand the order parameter around
$\phi_0 = \sqrt{b/c}$,
$\phi({\bf x},t) = \phi_0 + \tilde{\phi} ({\bf x},t)$.
Substituting this into the causal CH equation and
ignoring the non-linear terms, we obtain an equation for
the {}Fourier transformed fluctuations, $\tilde{\phi}({\bf k},t)$,
analogous to (\ref{shorttime}),
\begin{eqnarray}
\!\! \gamma \frac{\partial^2}{\partial t^2} \tilde{\phi} ({\bf k},t) +
\frac{\partial }{\partial t}\tilde{\phi}({\bf k},t)
\! \approx \! -\Gamma {\bf k}^2 ( a {\bf k}^2 + 2b )
\tilde{\phi} ({\bf k},t),
\label{eq longtime}
\end{eqnarray}
whose solution is
\begin{eqnarray}
\tilde{\phi}_c ({\bf k}, t) =
A_{\bf k} e^{\lambda'_+ t} + B_{\bf k}
e^{\lambda'_- t} \;,
\label{longtime}
\end{eqnarray}
where $A_{\bf k}$ and $B_{\bf k}$ are arbitrary constants and
$\lambda'_{\pm} = (-1 \pm \sqrt{1 - 4 \gamma \Gamma {\bf k}^2 (
a {\bf k}^2 +2 b)} )/(2\gamma)$.
On the other hand, the solution for $\tilde{\phi}_{nc}({\bf k},t)$,
coming from the noncausal CH equation, is
\begin{eqnarray}
\tilde{\phi}_{nc}({\bf k},t)= C_{\bf k} e^{-  \Gamma {\bf k}^2 (
2b + a {\bf k}^2) t}. \label{longtime nc}
\end{eqnarray}
{}From Eq.~(\ref{longtime}) it follows that there is a critical
momentum 
${k'}^2_c = [-b + \sqrt{b^2 + a/(4\gamma \Gamma)}]/a$, 
such that for $k \equiv |{\bf k}|< {k^\prime}_c$, the relaxation 
time is $1/(-\lambda'_+)$ and the long time fluctuation modes are 
overdamped, while for $k > {k^\prime}_c$ the relaxation time is 
$2\gamma$ and the relaxation is accompanied by oscillatory 
fluctuation modes.
These features are not exhibited by the fluctuation mode
solution coming from the noncausal CH equation,
Eq. (\ref{longtime nc}), which is always overdamped for any momentum $k$.
Now, from the time {}Fourier transform of Eq. (\ref{eq longtime}), we
obtain the expression for the frequency in terms of the
wave-number (neglecting the complex part coming from the first
order derivative in time), $\omega({\bf k}) = \sqrt{\Gamma {\bf
k}^2 (a {\bf k}^2 + 2 b)/\gamma}$.
This defines the wave-number velocity for the fluctuations.
Calculating it at the critical value ${{\bf k}'}^2_c$,
which  characterizes the maximum scale in the momentum space
(because the higher momentum modes
have rapid oscillations and cancel in computing averages),
the maximum wave-number velocity is found to be given by
\begin{equation}
v^2({\bf k}'_c) = 
\frac{2 D}{\gamma} \frac{1 + \xi^2/(2\gamma D)}{1+
\sqrt{1 + \xi^2/(2\gamma D) }}\;,
\label{vk}
\end{equation}

\noindent 
where we have defined $D=2 b \Gamma$ and the correlation
length $\xi = \sqrt{a/b}$. Taking the limit $\gamma \to 0$ in
Eq.~(\ref{vk}), one has that $v^2({\bf k}'_c)$ goes to infinity,
which is consistent with our previous assertion that the original
CH equation violates causality. Taking the opposite limit of large
values for~$\gamma$, Eq. (\ref{vk}) gives a constraint condition
relating the parameters $D$ and $\gamma$, $\gamma > D$, for which the
wave-number velocity should have limiting value one, which leads
to a causal propagation. 
{}For values of $\gamma$ satisfying the constraint $\gamma >D$ 
we can always find
allowed values of parameters for which Eq. (\ref{causalCH}) is
causal. This explains our previous notations. 

Let us see how this also applies for instance to the spinodal modes
and then use these results e.g. for the problem of
searching a signal of a phase transition in RHIC.
By considering the fastest-growing mode of the SD,
which is defined by (using Eq. (\ref{phi-c}))
\begin{eqnarray}
\left. \frac{\partial}{\partial k}
\frac{-1+\lambda}{2\gamma}\right|_{k^c_r} = 0,
\end{eqnarray}
we derive the time scale of the fastest mode to be
\begin{eqnarray}
\tau_c(k_r) = \frac{2\gamma}{-1+\sqrt{1+\gamma D/(2\xi^2)}}.
\label{tauc}
\end{eqnarray}
When $\gamma$ is very small, the time scale is reduced to
$\tau_{nc}(k_r) = 8\xi^2/D$, which agrees with the time scale of
the SD without the effect of memory~\cite{gavin2}. Since the modes
that give the result of Eq.~(\ref{tauc}) are in fact the dominant
spinodal modes and $\tau_{nc} < \tau_{c}$, this reflects itself in
an overall delay of the time formation of domains, as described by
the causal CH equation compared to the noncausal one, as the phase
transition proceeds. This feature is confirmed by our simulations
shown below. {}For a problem like RHIC and a possible signature of
a phase transition coming from it, this difference in time scales
can be very pronounced and lead to a striking effect that a
signal, like charge fluctuations and domain formation, can be so
much delayed that possibly could not be observed in the current
experiments. {}For instance, in RHIC, the correlation length is
typically $\xi \sim 1$ fm. Using also the
relation between the parameters $D$ and $\gamma$ obtained for
a quark plasma \cite{Greiner}, $\gamma = 3 D$, which is consistent 
with our constraint condition obtained from Eq. (\ref{vk}),  
and considering $D \sim 3.7 $~fm
\cite{gavin2,gavin}, we obtain from Eq.~(\ref{tauc}) that
$\tau_c\sim 6.1$ fm, which is to be compared with the result
$\tau_{nc}\sim 2.2$ fm. This represents almost a $200 \%$
difference for the time scales for the starting of the growth of
fluctuations in Eq.~(\ref{causalCH}) as compared to the ordinary
CH equation (for $\gamma=0$).

\begin{figure}[htb]
\centerline{\epsfig{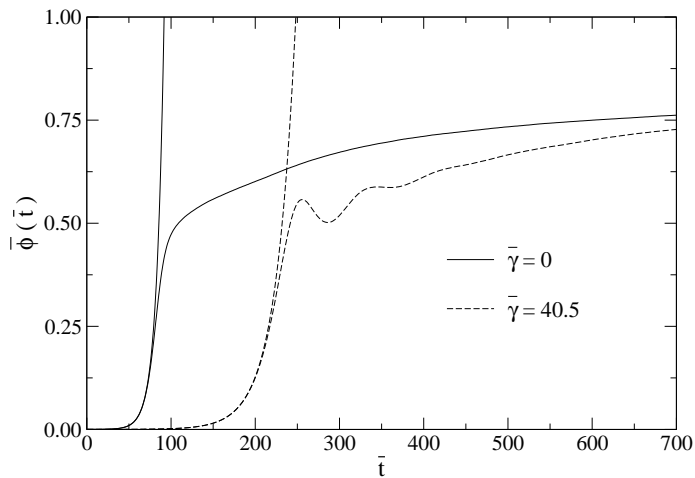}}
\caption{Volume average of $\bar\phi(\bar t)$ as a
function of dimensionless time $\bar t$ for the noncausal (solid)
and causal (dashed) solutions. The two curves that exit the top of 
the plot are results from the linearized theory.} \label{fig:phi}
\end{figure}

We have solved Eq.~(\ref{causalCH}) numerically on a discrete
spatial square lattice  using a semi-implicit scheme in time, with
a fast {}Fourier transform in the spatial coordinates~\cite{CS}.
We have checked the stability of the results by changing lattice
spacings and time steps. In addition, for $\gamma > 1$ we have 
also used a leap-frog algorithm and the results
obtained with both methods agreed very well. {}For the noncausal
$\gamma = 0$ equation, we used as initial condition $\phi({\bf
x},t=0)$ a random distribution in space with zero average and
amplitude $10^{-3}$. {}For $\gamma \neq 0$, we used in addition
the condition that at $t = 0$ the first-order derivative of $\phi$
is zero. {}For the numerical work, Eq. (\ref{causalCH}) is
re-parameterized to dimensionless variables, conveniently defined
by time $\bar{t} = (8/\tau_{nc}) t= (2 b^2 \Gamma/a) t$, space
coordinates $\bar{x}_i = x_i/\xi$, field $\bar{\phi} = \sqrt{c/b}
\, \phi$ and $\bar{\gamma} = (8/\tau_{nc})\gamma$. In terms of
these variables Eq. (\ref{causalCH}) becomes function of only one
parameter, $\bar{\gamma}$. Eq. (\ref{causalCH}) was next solved
for several values of $\bar{\gamma}$. Two representatives results,
for $\bar{\gamma}=0$ and $\bar{\gamma} = 40.5$ (for the example
analyzed in the previous paragraph), are shown in
{}Fig.~\ref{fig:phi}.

\vspace{0.64cm}

In {}Fig.~\ref{fig:phi} we present the time evolution of the
volume average of the order parameter $\phi$, defined as $\bar
\phi(t) = \frac{1}{N^3} \sum_{{\bf x} } \langle \bar \phi({\bf
x},t)_+ \rangle$, where $N^3$ is the total number of lattice
points and the average is taken over different initial random
configurations. $\phi({\bf x},t)_+$ indicates that only the
positive values of the field are considered (i.e., a specific
direction for the field has been selected). The rapid increase of
$\bar{\phi}$ reflects the phenomenon of SD. The figure also
shows the results by solving the linear equation. It shows that it
performs extremely well up to and right after the spinodal growth
of the order parameter, then justifying our previous analytical
results based on the solution of the linear equation. The effect
of $\gamma$ is seen to be more important at earlier times,
consistent with the memory function used and becomes less
important after the rapid growth of the order parameter
(the spinodal explosion). It also shows the effect
of a finite $\gamma$, increasing dramatically the delay of the
spinodal explosion, as predicted by our previous analytical
results. The time for reaching equilibrium is seen to be very
long, as is common with the traditional noncausal CH equation.
Also apparent from {}Fig.~\ref{fig:phi} are the oscillations in
the order parameter for a finite $\gamma$, also predicted by our
previous analysis. This is due to the increasing importance of the
second-order time derivative as compared to the first-order one as
$\gamma$ increases, i.e. as $\gamma$ increases the dissipation
term becomes less important and the equation becomes more and more
a  wave-like equation. The estimated delay for the thermalization
is even larger than the recent estimation~\cite{FK} for the time
delay of the relaxation of a nonconserved order parameter.

\begin{figure}[th]
\vspace{0.64cm}
\centerline{\epsfig{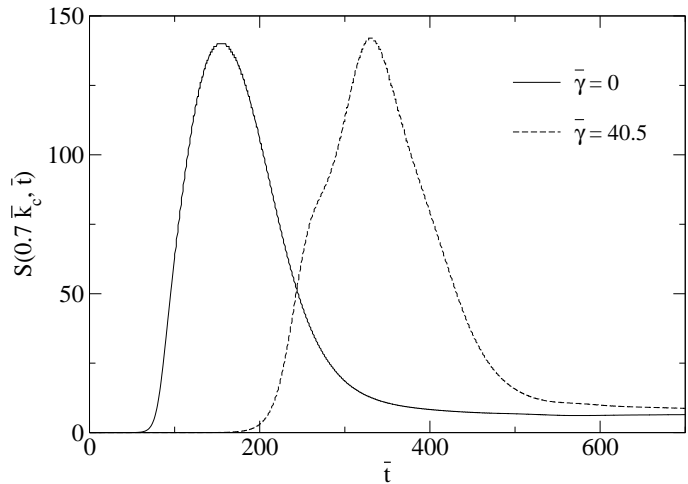}}
\caption{Spherically averaged structure function
as a function of dimensionless time $\bar t$ for a $\bar k = 0.7
\bar k_c$ for the noncausal (solid) and causal (dashed)
solutions.} \label{fig:struct}
\end{figure}

We have also investigated the effect of memory for the
structure function, $S(k,t) = |\phi({\bf k},t)|^2$.  This quantity is
important because it provides information on the space-time
coarsening of the domains of the different phases. In
{}Fig.~\ref{fig:struct} we present the results for the spherically
averaged value of $S$ for $k = 0.7 k_c$ (to emphasize the fast
growth of the long wavelength fluctuations with $k < k_c$). The
spherical averaging for a given $k$ was done over momenta $k_r =
\sqrt{k^2_x + k^2_y + k^2_z}$ such that $k - 0.1 \Delta \leq k_r
\leq k + 0.1 \Delta$, with $\Delta = 2\pi/L$, where $L$ is the size
of the lattice. Consistently with Fig.~\ref{fig:phi}, this figure
shows the dramatic delay for the spinodal growth.

As conclusion, we have introduced memory into the CH
equation. For a physical situation typical for the phenomenology
of RHIC, we found that these ME
can delay substantially the phase-separation process and
consequently, there might not be enough time for the system to
thermalize before the breakdown of the system due to expansion.

\acknowledgments

T.K. is grateful to Y. Hama for drawing attention to
Ref.~\cite{Hama}. The authors also thank CNPq, FAPERJ and FAPESP
for financial support.

\end{document}